\documentclass[conference]{IEEEtran}
\IEEEoverridecommandlockouts

\usepackage[utf8]{inputenc}
\usepackage{cite}
\usepackage{amsmath,amssymb,amsfonts}
\usepackage{algorithmic}
\usepackage{graphicx}
\usepackage{textcomp}
\usepackage{xcolor}
\usepackage[nolist, printonlyused, nohyperlinks]{acronym}
\usepackage{paralist}
\usepackage{enumitem}
\usepackage{wrapfig}
\usepackage{placeins}

\usepackage{subfigure}

\usepackage{amsmath}
\usepackage{amssymb}
\usepackage{float}

\usepackage{multicol}
\usepackage{xcolor}
\usepackage[most]{tcolorbox}
\usepackage{blindtext}
\usepackage{graphicx}
\graphicspath{{figures/}}
\usepackage{mathtools}
\usepackage{tikz}
\usepackage{pgfplots}
\usepackage{pgfplotstable}
\usepackage{wrapfig}
\pgfplotsset{compat=1.13}

\usepackage[colorinlistoftodos,prependcaption,textsize=tiny]{todonotes}
\presetkeys{todonotes}{inline,backgroundcolor=yellow}{}

\usepackage{lipsum}

\usepackage{tabularx}
\usepackage[multiple]{footmisc}

\usepackage{siunitx}

\usepackage{array}
\newcolumntype{P}[1]{>{\centering\arraybackslash}p{#1}}
\usepackage{threeparttable}

\def\BibTeX{{\rm B\kern-.05em{\sc i\kern-.025em b}\kern-.08em
    T\kern-.1667em\lower.7ex\hbox{E}\kern-.125emX}}
\begin{document}

\title{Problems with Risk Matrices Using Ordinal Scales}

\author{
\IEEEauthorblockN{Michael Krisper}
\IEEEauthorblockA{\textit{Institute of Technical Informatics} \\
\textit{Graz University of Technology}\\
Graz, Austria \\
michael.krisper@tugraz.at}
}

\maketitle

\def\BibTeX{{\rm B\kern-.05em{\sc i\kern-.025em b}\kern-.08em
    T\kern-.1667em\lower.7ex\hbox{E}\kern-.125emX}}

\newcommand{\mv}[1]{\mathit{#1}}
\newcommand{\mo}[1]{\operatorname{#1}}

\newcommand{\includevisio}[4]{\includegraphics[clip, trim=#1 #1 #1 #1, #2]{#3}\label{#4}}
\newcommand{\includevisiocap}[5]{\includegraphics[clip, trim=#1 #1 #1 #1, #2]{#3}\caption{#5}\label{#4}}
\newcommand{\includevisionolabel}[3]{\includegraphics[clip, trim=#1 #1 #1 #1, #2]{#3}}

\newcommand{\myfig}[1]{Figure~\ref{#1}}
\newcommand{\mytable}[1]{Table~\ref{#1}}
\newcommand{\mysec}[1]{Section~\ref{#1}}
\newcommand{\mycha}[1]{Chapter~\ref{#1}}

\renewcommand{\baselinestretch}{1.0}

\newenvironment{myblock}[1]{%
    \tcolorbox[beamer,%
    noparskip,breakable,%
    colback=LightBlue,colframe=DarkBlue,%
    colbacklower=DarkBlue!75!LightBlue,%
    title=#1]}%
    {\endtcolorbox}

\newcommand{\Rplus}{\protect\hspace{-.1em}\protect\raisebox{.35ex}{\smaller{\smaller\textbf{+}}}}
\newcommand{\Cpp}{\mbox{C\Rplus\Rplus}\xspace}

\newcommand{\myhl}[1]{\textsc{#1}}
\newcommand{\myhla}[1]{\textit{#1}}
\newcommand{\myhlp}[1]{\uppercase{#1}}

\newenvironment{benumerate}[1]{
    \let\oldItem\item
    \def\item{\addtocounter{enumi}{-2}\oldItem}
    \begin{enumerate}
    \setcounter{enumi}{#1}
    \addtocounter{enumi}{1}
}{
    \end{enumerate}
}

\begin{acronym}
 \acro{ADAS}{advanced driver assistant system}
 \acro{CD}{continuous deployment}
 \acro{CI}{continuous integration}
 \acro{CPS}{cyber-physical system}
 \acro{CRUD}{create-read-update-delete}
 \acro{IIoT}{industrial Internet-of-Things}
 \acro{IoT}{Internet-of-Things}
 \acro{IE}{industrial Ethernet}
 \acro{IT}{information technology}
 \acro{M2M}{machine-to-machine}
 \acro{OCC}{operations control center}
 \acro{OSI}{open systems interconnection}
 \acro{OT}{operation technology}
 \acro{PLC}{programmable logic controller}
 \acro{QoS}{quality of service}
 \acro{RT}{real time}
 \acro{S2S}{service-to-service}
 \acro{SCADA}{supervisory control and data acquisition}
 \acro{SOA}{service-oriented architecture}
 \acro{SoS}{system of systems}
 \acro{SPoF}{single point of failure}

 \acro{ATA}{attack tree analysis}
 \acro{2PC}{two-phase commit}
 \acro{ACID}{atomicity-consistency-isolation-durability}
 \acro{BDMP}{boolean logic driven markov processes}
 \acro{BRA}{binary risk analysis}
 \acro{CCP}{common closure principle}
 \acro{DDD}{domain-driven design}
 \acro{DDS}{data distribution service}
 \acro{DGDS}{distributed global data space}
 \acro{DSS}{domain specific service}
 \acro{E/E/PE}{electric/electronic/programmable electronic}
 \acro{FIT}{failures in time}
 \acro{FMEA}{failure mode and effects analysis}
 \acro{FMVEA}{failure mode, vulnerabilities and effects analysis}
 \acro{FTA}{fault tree analysis}
 \acro{GIG}{global-information-grid}
 \acro{HARA}{hazard analysis and risk assessment}
 \acro{OMG}{object management group}
 \acro{OOD}{object-oriented design}
 \acro{REST}{representational state transfer}
 \acro{RPN}{risk priority number}
 \acro{RTE}{real time ethernet}
 \acro{SAHARA}{security-aware hazard and risk analysis}
 \acro{SecL}{security level}
 \acro{SRP}{single responsibility principle}
 \acro{TSN}{time-sensitive networking}
 \acro{FAIR}{factor analysis of information risk}
 \acro{LEF}{loss event frequency}
 \acro{LM}{loss magnitude}
 \acro{TEF}{thread event frequency}
 \acro{Vuln}{vulnerability}
 \acro{TCap}{thread capability}
 \acro{RS}{resistance strength}
\end{acronym}




\begin{abstract}
In this paper, we discuss various problems in the usage and definition of risk matrices. We give an overview of the general process of risk assessment with risk matrices and ordinal scales. Furthermore, we explain the fallacies in each phase of this process and give hints on which decisions may lead to more problems than others and how to avoid them. Among those 24 discussed problems are ordinal scales, semi-quantitative arithmetics, range compression, risk inversion, ambiguity, and neglection of uncertainty. Finally, we make a case for avoiding risk matrices altogether and instead propose using fully quantitative risk assessment methods.
\end{abstract}


\section{Introduction}

Risk matrices are established tools to assess and rank risks in many domains and industries. They have become so common that everyone accepts and uses them without question. They have many seemingly benefits like the simplicity of usage, different coloring systems with traffic light semantics, intuitive understanding, and are seemingly proven in use over many decades. When there is little or no data available, they are praised as the weapon of choice for tackling risks and estimates in projects. Nevertheless, they have many flaws and problems that will be covered in this work.

More than a decade ago, Anthony Cox and his team started a riot against risk matrices which has not come to an end since \cite{cox_limitations_2005,cox_limitations_2007,cox_whats_2008,cox_risk_2009,cox_confronting_2012}. He has shown and proven that risk matrices have severe problems that could diminish their usefulness to the point where they become even worse than random. More and more scientists, engineers, and managers have since supported the cause against risk matrices, and amongst them, Douglas Hubbard is one of the most prominent ones. In his excellent book series ``How to measure anything'' \cite{hubbard_how_2014,hubbard_how_2016}, he also defends Cox et al. and demonstrates some ideas and techniques for a quantitative risk assessment method to overcome and avoid the problems of classical risk matrices.

In this work we build upon the findings of Cox \cite{cox_whats_2008}, Hubbard \cite{hubbard_how_2016}, Artzner \cite{artzner_coherent_1999}, Talbot \cite{julian_talbot_whats_2018}, Kahnemann \& Tversky \cite{kahneman_subjective_1972}, as well as many others over the last few decades. We show that risk matrices today still have some flaws, fallacies, and pitfalls and explain what those are. By showing them, we want to, once more, state a case for fully quantitative risk assessment using quantitative value ranges, ratio scales, and probability distributions, which are considering the uncertainties throughout the risk analysis. Our focus in this paper lies in summarizing pitfalls and fallacies in risk assessment using ordinal scales and risk matrices with concise and understandable explanations and examples.

The paper is structured as follows: After this introduction, the motivation sections show some examples of what can go wrong using risk matrices and its consequence. Subsequently, we directly dive into the overview and descriptions of the problems, pitfalls, and fallacies of risk matrices found in the literature.

The authors are researchers at the Graz University of Technology and have a background in automotive safety, quality, and security. This work aims to show the problems of qualitative risk assessment methods to argue towards quantitative methods. In particular, in our research group, we are currently working on a method for integrated quantitative risk assessment \cite{dobaj_towards_2019}, which combines safety and security. For that, we are developing a tool based on attack-trees using truly quantitative methods called RISKEE \cite{krisper_riskee:_2019}.

\section{Motivation}

\textit{``What is so bad about risk matrices?''}, one may ask, \textit{``they are so widely accepted and established tools, they cannot be wrong.''}. Only because something is established does not mean it is without any flaws. In this section, we show some examples of pitfalls that may occur when using risk matrices. We will show some artificial examples and real-life use-cases, where the ranking of risks with risk matrices is illogical, unreasonable, or leads to problems.

\subsection{Oil Leakage and the MIL-STD882C}

\begin{figure}
  \centering
  \includegraphics[width=\linewidth]{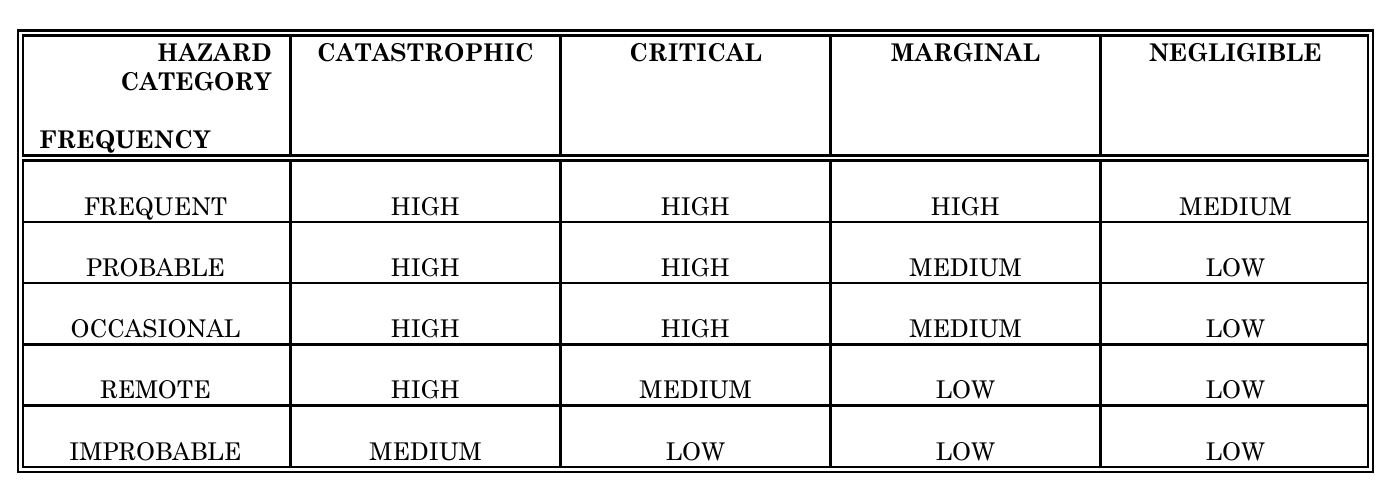}
  \caption{An example of a risk matrix defined in the standard MIL-STD-882C \cite{882C}.}
  \label{fig:risk-matrix-882c}
\end{figure}

This example was taken from \emph{``What's wrong with risk matrices''} by Cox et al. \cite{cox_whats_2008} and shows the problem of risk inversion (see the problem (U)). In this example, two physical hazards for environmental damage (fuel leakage in this case) are compared. The first event consists of 1 ounce of fuel spills five times per hour. The second event causes more damage but happens less frequently, with 10 pounds of fuel leaking once per week. According to the military standard, 882C \cite{882C}, both would arguably get the highest frequency rating, but the one leaking 1 ounce would get a negligible hazard rating (resulting in a MEDIUM score). In contrast, the 10-pound event would get a marginal or even critical severity (resulting in a HIGH score). If we compute the risks quantitatively, we get another result: The 1-ounce event produces 52.5 pounds of leakage per week (1oz*5*24*7), while the 10-pound event leaks 10 pounds. Thus, the first event should be rated way higher than the second event, which it is not. Figure~\ref{fig:risk-matrix-882c} shows the risk matrix taken from MIL-STD-882C. This example shows cases where the qualitative risk score does not reflect the actual quantitative risk. Even worse, it results in an inverse order for the events' priorities, which is the consequence of risk inversion.

\subsection{Failure Mode and Effects Analysis (FMEA)}

Another example is the risk matrix in the Failure Mode and Effects Analysis (FMEA) \cite{iec60812_2006}. Especially here, the severity scale is problematic because it combines four different effects scales into one ordinal scale and assigns them ranks from 1 to 10 (annoyances, failure of secondary functions, failure of primary functions, failure of safety). Furthermore, this ordinal scale is multiplied with other influence factors to get a resulting risk priority number (RPN), although multiplication is not defined on ordinal scales. Furthermore, a higher detectability factor could reduce the RPN tremendously but does not reduce the actual hazard. Is the risk of a hazard less severe just because we can detect it?

\subsection{Hazard and Risk Analysis (HARA)}
\begin{figure}
  \centering
  \includegraphics[width=\linewidth]{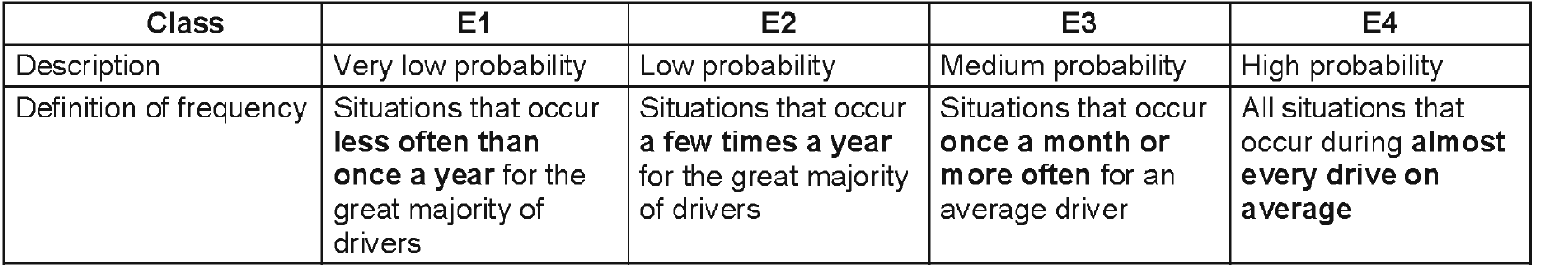}
  \caption{Informative examples for the exposure in the ISO 26262 \cite{iso_26262}.}
  \label{fig:hara-exposure}
\end{figure}

The Hazard and Risk Analysis (HARA) \cite{iso_26262} is done at the concept and system level in the early stages of product development. The problem of this method is the ambiguity of the input scales, in particular, exposure. First, let us summarize the method itself: During the analysis, one assesses possible hazardous scenarios for their severity, exposure, and controllability. All three values are logarithmically distributed ordinal scales that assign a number for the rank. The ranks for the individual scales get added up, and the resulting number is translated into an ASIL (automotive safety integrity level) classification. Depending on the ASIL, there are exponentially more complex requirements to fulfill for developing a product in a safe way. These requirements become so high that it is challenging to implement them using only a single component. Thus, the ASIL can be decomposed into subsystems with lower ASIL but have to be independent, redundant, and diverse to avoid common cause failures. This decomposition increases the costs tremendously. A false ranking of the initial values has severe consequences to all subsequent development efforts and costs of a product. Especially border cases are the problem here: Decreasing the score of a borderline case could decrease the resulting ASIL from D to C, which cuts the effort for product development to half. Let us examine this in the case of the exposure score. The exposure can be defined in two ways: either via the frequency of occurrence over time or as the proportion of duration in hazard situations compared to the total operating time of a product. These two aspects are reasonable because sometimes the frequency is needed(e.g., traffic situations), and sometimes the event's duration (e.g., radiation exposures). Here, ambiguity strikes the hardest: Changing the argumentation from one to the other could change the score entirely.
Furthermore, even when staying in the same category, the scales are ambiguous. Figure \ref{fig:hara-exposure} shows an excerpt for the exposure from the informative annex of the ISO 26262 \cite{iso_26262}. Exposure rank 2 states \emph{a few times per year for most drivers}, while E3 states \emph{once a month or more often for an average driver}. There are two ambiguities here: Firstly, what exactly is the definition of the majority of drivers and the average driver? Does the \emph{majority of drivers} mean more than 80\%, 90\%, 99\%, or is it 51\%? Secondly, where is the border between a few times per year and once a month or more often? Is it six times per year? Even if the boundaries would have been defined exactly, a quantification problem is still left, which we will discuss later on (problem (J)).

\begin{figure}
  \centering
  \includegraphics[width=\linewidth]{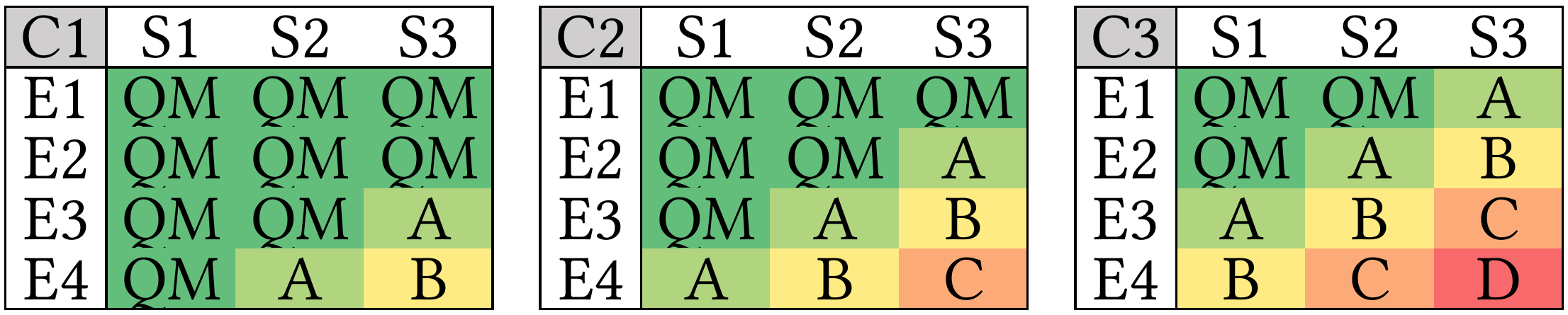}
  \caption{The risk matrices for the Hazard and Risk analysis in ISO 26262 \cite{iso_26262}, illustrated here by splitting it up into three parts with different controllability scores.}
  \label{fig:hara-side-by-side}
\end{figure}

\section{The Problems of risk matrices}

In the following sections, we go over the typical process of risk scoring methods based on risk matrices and explain why this is problematic and what the problems are. Furthermore, we compare this to quantitative risk assessment methods to show that they do not suffer from the described problems and should always be preferred over qualitative methods.

\begin{figure*}
  \centering
  \includegraphics[width=\linewidth]{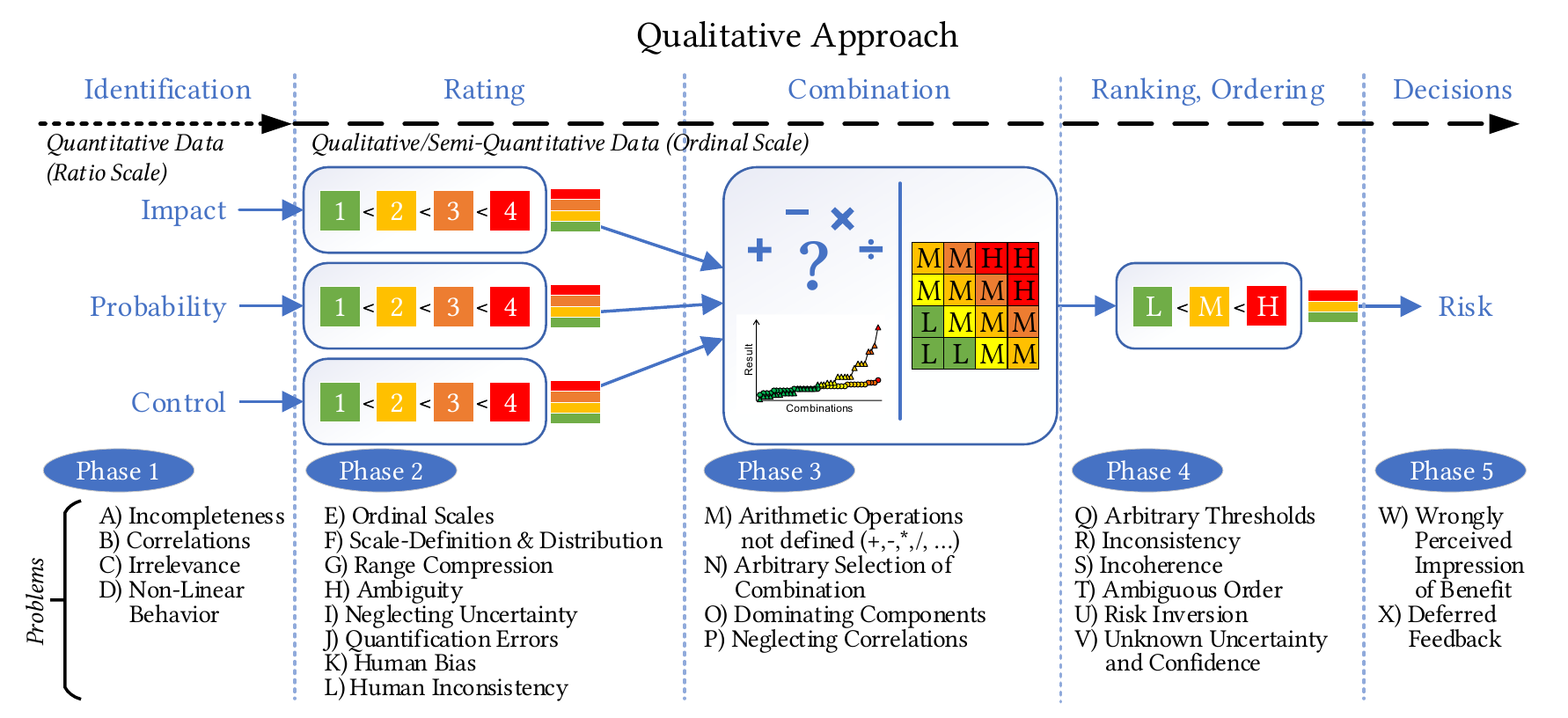}
  \caption{The flow of information and the processing phases during a typical qualitative risk asessement approach based on ordinal scales and risk matrices.}
  \label{fig:overview-qualitative}
\end{figure*}

\begin{figure*}
  \centering
  \includegraphics[width=\linewidth]{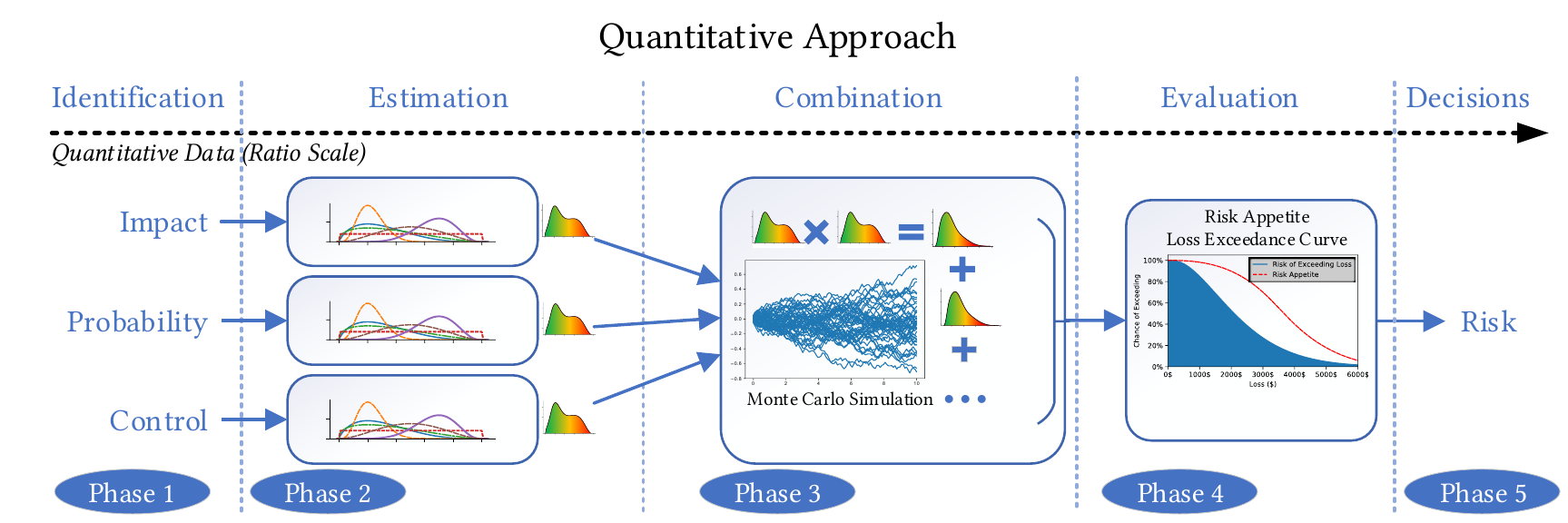}
  \caption{The flow of information in a quantitative risk assessment approach using ratio scales, probability distributions, monte carlo simulation, and loss exceedance curves.}
  \label{fig:overview-quantitative}
\end{figure*}

Qualitative (or semi-quantitative) risk assessment methods based on ordinal scales and risk matrices typically are done in five phases, which is illustrated by Figure~\ref{fig:overview-qualitative}. It shows an overview of the five phases and enlists the problems that may occur in each phase. Just to give a comparison, Figure~\ref{fig:overview-quantitative} shows the corresponding quantitative approach, which also has five phases. The risk score is computed quantitatively by estimating plausible ranges of input factors, simulating them using Monte-Carlo simulation, and comparing them to the risk appetite (also called risk affinity) using a loss exceedance curve \cite{krisper_riskee:_2019,hubbard_how_2016}. 

Nevertheless, in the following sections, we will discuss the five phases of qualitative risk assessment and their problems and compare them to the quantitative approach whenever reasonable:
\begin{itemize}
    \item \emph{\textbf{Phase 1: Identifying the influence factors}}. First, a set of influential factors has to be defined.
    \item \emph{\textbf{Phase 2: Rating of the factors}}. In this phase, the input values are rated according to some scale.
    \item \emph{\textbf{Phase 3: Combining the ratings}}. The scorings are combined to get a final risk score.
    \item \emph{\textbf{Phase 4: Ranking the combinations}}. The risk score is ranked against other risks or filtered by some threshold.
    \item \emph{\textbf{Phase 5: Decision making based on the rankings}}. In the end, decisions have to be made which risks to reduce. The goal is to bring the risks down to a tolerable level.
\end{itemize}


\subsection*{\textbf{Phase 1: Identifying the influence factors}}
Before doing any risk assessment, one has to define the influence factors that affect the system's risk, which is assessed. Most standardized risk assessment methods defined the influence factors right away, e.g., impact, probability, severity, exposure, utility, loss. Some leave it open to be defined by the practitioners. Others are only defined within a single organization to be specialized for a specific situation, e.g., for actuary sciences, insurances, medicine, or the financial sector. For such industries, the identification of influence factors plays a massive role in risk assessment.
The number of influence factors often is related to the used method. General risk assessment methods tend to use only two or three factors; specialized ones tend to use more. Furthermore, multiplicative methods also tend to use lesser factors, like two or three, and additive ones use more in general. Examples of the most frequently used factors for general risk assessment methods are the following:
\begin{itemize}
    \item \emph{Impact}: The impact corresponds to the actual outcome when the risk event occurs. A higher impact also means higher risk. This is also called severity, loss, magnitude, harm, effect, threat, consequence, or utility.
    \item \emph{Probability}: The probability defines some notion of the likelihood that an event occurs. A higher probability also means higher risk. This is often also called exposure, likelihood, vulnerability, frequency
    \item \emph{Control}: The control factor corresponds to a risk reduction possibility. Higher control over a situation results in lowering risk. Often this is also called controllability, mitigation, reduction, protection, detection, or reaction. Many methods omit this factor by reasoning that its behavior can also be modeled by reducing the probability or impact.
\end{itemize}

In contrast to that, many additive methods tend to use more specific factors like demographic features, medical attributes, lifestyle attributes. A recently highly discussed example in Austria was the introduction of a rating scheme for the unemployment office in 2018 (Arbeitsmarktservice, AMS) \cite{ams-algorithmus}. This was a weighted additive scheme for rating the risk of future unemployment (or otherwise put: the chance for employability). It was based on several demographic and social factors, including, e.g., gender, age, disability, career, education, and many more. The weights for combining the factors were inferred via historical data and statistics and reflected society's bias explicitly. For example, in that scoring algorithm, women have lower job chances than men. This resulted in public discussions, similar to that for amazon's automatic firing algorithm \cite{lecher_how_2019}. More about that subject will be discussed in phase 3: \emph{Combining the ratings}.

Now we discuss the problems which may occur in this first phase of selecting the influence factors:

\paragraph*{\textbf{(A) Incompleteness}}
    To do a useful risk assessment, all essential factors have to be accounted for in the analysis. However, sometimes, factors are forgotten or overlooked. This could happen unknowingly or due to ignorance or inexperience. It could be complicated to determine the significant factors that influence the risk, especially for complex behavioral or technical systems. This is not only a problem for qualitative approaches but may happen in quantitative approaches also.

\paragraph*{\textbf{(B) Correlations}}
    The selected factors could be correlated to each other, or in other words: they could influence each other. Even more dangerous is a negative correlation: If one factor grows, others may decline. This could result in worse than random results \cite{cox_whats_2008}. There is also a common pitfall of correlated influence factors regarding the view of information theory: When they are strongly correlated, they do not deliver more information than one of them alone. If two scales would always show the same value, it would suffice to use just one because the information gain would be the same. So the solution to this would be to try to avoid correlated factors. This is hard to detect for qualitative approaches, but for quantitative approaches, this could be detected via statistical methods (see problem P for more information).

\paragraph*{\textbf{(C) Irrelevance}}
    Irrelevant factors make the risk assessment more difficult because they have to judge, evaluated, and discussed but have no real impact on the result. There are two aspects of this: The first aspect is real irrelevance - a factor does not increase or decrease the resulting risk. Then it can be skipped in the analysis. The second aspect is, if a factor is the same for all risks, it has no relevance anymore. For example, if we would use a priority factor for risks, and every risk would be rated as a high priority, this does not help the final ranking because every risk would be equal.

\paragraph*{\textbf{(D) Nonlinear Behavior}}
    An input factor could have non-linear behavior, making it difficult to model or rate in the next phase. Logarithmic or polynomial scales could cope with this. However, the more complex a factor is, the more difficult it is to model and judge, e.g., the driving speed as a risk factor is perceived as linear. However, the actual risk increases quadratically or even exponentially \cite{martin_pedestrian_2017,jurewicz_exploration_2016,aarts_driving_2006}.

The mentioned problems apply to qualitative and quantitative risk assessment methods. Nevertheless, quantitative methods can at least tackle them by using mathematical tools. With statistical sensitivity analysis, correlations and irrelevances can be detected (ANOVA, Correlation Coefficients, Hypothesis Tests). Furthermore, non-linear behavior can be modeled in mathematical equations in quantitative models, which would be difficult in qualitative ordinal scales. Only incompleteness is a problem that is hard to solve for both methods. It is not trivial to detect that a factor is missing, which boils down to the often-cited management mantra by Tom DeMarco: \emph{``You cannot control what you cannot measure''} \cite{demarco_controlling_1982}. Nevertheless, Hubbard et al. propose regular and immediate feedback as a tool to evaluate risk assessment methods \cite{hubbard_how_2016}. In such a way, it could be detected that a model is not realistic and may have left out some crucial influence factors.

\subsection*{\textbf{Phase 2: Rating of the influence factors}}
After the influence factors are identified, they have to be estimated and rated. This is typically done using an ordinal scale defined by the used method or standard or has some internal company or domain-specific definition. In either way, it comes down to deciding for a class on an ordinal scale which the factor corresponds to in order to be able to rank the factors and use them for later risk comparison. In quantitative methods, this is done differently: here, an actual value, range, or even distribution on a ratio scale, which represents the reality, is chosen (no classification, just estimation of real values including the respective uncertainty). The assignment to classes in qualitative methods is one of the most discussed problem areas in literature. We will go through the problems and show how quantitative methods can cope with most of them:

\paragraph*{\textbf{(E) Ordinal Scales}}
Qualitative Risk assessment methods mostly use ordinal scales. According to Stevens \cite{stevens_theoryof_1946}, ordinal scales only allow for ordering or ranking the items. Therefore arithmetic operations like addition or multiplication are undefined. Nevertheless, in risk assessment methods, this is done nearly all the time without question. Stevens defines the following scales, and their respective defined operations \cite{stevens_theoryof_1946}:

\begin{itemize}
    \item \textbf{Nominal scale}: Defines equality or inequality ($=, \neq$) of items. Examples: Different kinds of fruits like oranges, apples, or pears.
    \item \textbf{Ordinal scale}: Defines ordering relations ($<, >$) amongst items. Examples: School grades or the ranks in sports events.
    \item \textbf{Interval Scale}: Defines sums and differences ($+, -$) in addition to the ordering. Examples: Temperature; Time; often, values in sports events are stated as time differences compared to the first place.
    \item \textbf{Ratio scale}: Defines absolute ratios ($*, /$) between items in addition to the difference and ordering relations. Examples: Distance, Weight, Probabilities
\end{itemize}

A problem here is that by transforming quantitative values into a domain and scale, which only supports ordering relations, we lose the ability to do reasonable arithmetic, estimate uncertainty, or do any sophisticated mathematical analysis. Although the so-called ``semi-quantitative'' scales may give the illusion of doing calculations, the numbers are just placeholders for the class labels. They do not have mathematical foundations or actual connections to the real world. While one would refrain from multiplying ``words'' like \emph{high risk} and \emph{moderate impact} together, doing this with arbitrarily assigned numbers suddenly seems plausible. For example, if high risk=3 and moderate impact=3, then the risk is $3 \times 2 = 6$, but what is the meaning of 6?.

\paragraph*{\textbf{(F) Semi-Quantitative Scale-Definition}}
The problems begin with the definition of a semi-quantitative distribution on the ordinal scale. There are many articles on how to design a numeric ordinal scale for use in a semi-quantitative assessment e.g. \cite{knapp_treating_1990,ho_improving_2015,smith_risk_2009}. We give a short review of the different options here.    
What we want to achieve is a mapping from continuous quantitative data to a discrete ordinal scale. First, we have to decide how many ranks the ordinal scale should have and which ranges of values are assigned to which rank. Furthermore, if the ranks should be used for semi-quantitative arithmetics, the ranks must be assigned to numbers.

\begin{table*}
\caption{Different labels and semi-quantitative number assignments for ordinal ranks.}
\label{tab:scale-ranks}
\centering
\begin{tabular}{ccccccccccc}
Rating & Probability & Frequency & Increasing & Start at 0 & Decreasing & Centered & 3 Levels & 4 Levels & Spaced out & Exponential \\
\hline
Very Low  & Remotely & Never &1 & 0 & 5 & -2 &   & 1 & 2  & 1  \\
Low       & Unlikely     & Seldom & 2 & 1 & 4 & -1 & 1 & 2 & 4  & 2  \\
Medium    & As Likely as not & Sometimes &3 & 2 & 3 &  0 & 2 &   & 6  & 4  \\
High      & Likely       & Often &4 & 3 & 2 &  1 & 3 & 3 & 8  & 8  \\
Very High & Certain  & Always & 5 & 4 & 1 &  2 &   & 4 & 10 & 16 \\
\hline
\end{tabular}
\end{table*}

\emph{Decision 1: Number of Ranks}: Does the scale have three levels (e.g., high, medium, low), 4, 5, or even 10 or 100 levels? A high number leads to a seemingly continuous scale, while a lower number is more comfortable to judge due to its coarseness \cite{knapp_treating_1990}. An even number of levels has no neutral state, and therefore the assessment always points into a direction (either lower or higher). Uneven numbers of levels allow for a neutral position in the middle. In addition to that, Hubbard et al., as well as others, found out that people tend to avoid extreme positions \cite{hubbard_failure_2009, moors_effect_2014}. Therefore, it could sometimes be reasonable to add an even more extreme level to an existing scale to outwit the bias of avoiding the most extreme. Using increasing or decreasing numbers, and even how the scale is presented can affect the outcome \cite{moors_effect_2014}. A further aspect of this is the next decision is if every factor should have the same number of levels for simplicity's sake, or if they should have a different number of levels to fit the individual factor better. Scientists, like Rensis Likert, have researched the psychological effects of such scales for nearly a century now (he coined the term ``Likert-scale''). However, for conciseness, we leave out further psychological debate about psychometric scales and refer to \cite{malhotra_basic_2011,dawes_data_2008} for further information.
    
\begin{figure}
  \centering
  \includegraphics[width=\linewidth]{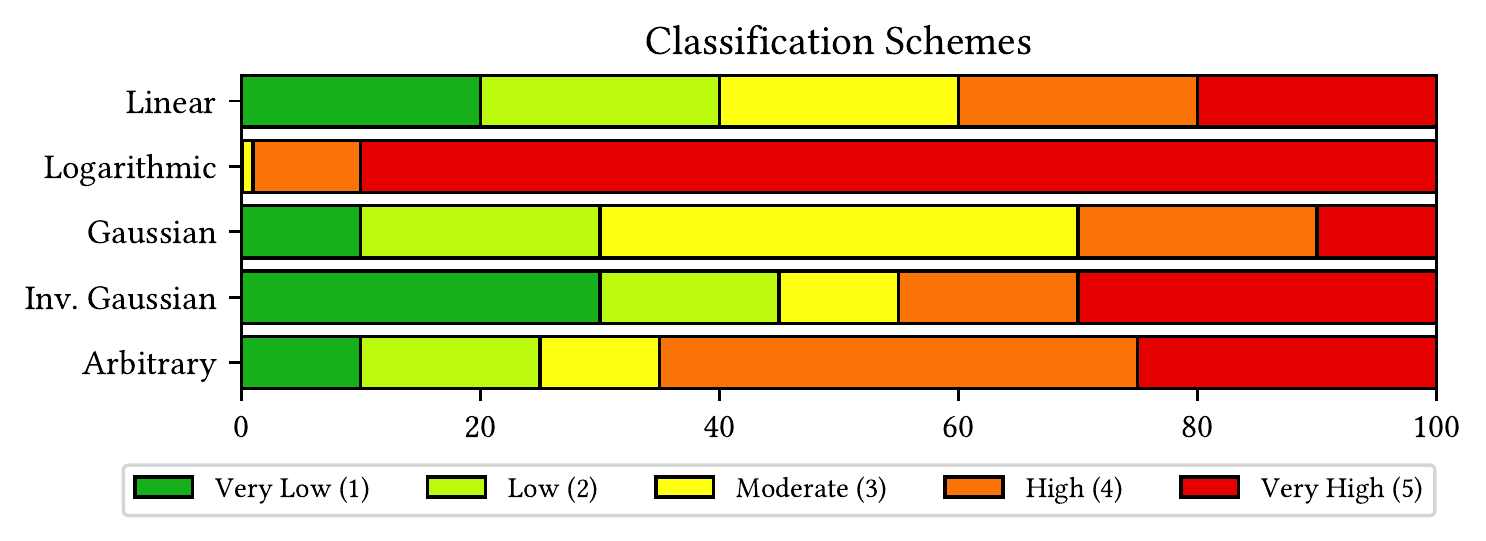}
  \caption{Illustration of different ranges for classification into ordinal scales.}
  \label{fig:distributions}
\end{figure}

\emph{Decision 2: Assignment of Quantitative Ranges to Ranks (Distribution)}: It is important to decide which ranges of values belong to which rank on the ordinal scale. Is this distribution scaled linear or logarithmic? Table~\ref{tab:scales} and Figure~\ref{fig:distributions} show different kinds of distribution numerically and graphically, and here we enlist and describe some of the most common ways to define the distribution of ranks:

\begin{itemize}
    
    \item \emph{Linear}: Linear-based scales split a value range into equally distributed ranges and assign labels to them. E.g., low, medium, high. Linear-based ordinal scales relate approximately to ratio scales but still have the problem of assigning arbitrary numbers to the value ranges, which dismisses all arithmetic semantics. Sometimes they are inappropriate because, in reality, processes often behave quadratic or even exponential, and we still want to be able to cover small differences for the lower values. A linear scale would have to be very big and unhandy to cope with such behavior (imagine a scale from 0 to 100 in 0.1-interval steps. It would have 1000 different levels, while a logarithmic scale would only have 4).
    
    \item \emph{Logarithmic}: a logarithmic scale considers processes covering large ranges while still being able to classify small ranges also, e.g., yearly, weekly, daily; small amounts of money vs. large amounts; 1, 10, 100, 1000; Injury is also very often scaled logarithmic (e.g., AIS scale).
    
    \item \emph{Normally Distributed (Gaussian)}: Scales that are arranged like a bell curve to distinguish between the tiny and huge exceptions, while average cases are all put into the same category. A variant of this is to arrange the values inversely, to distinguish the average cases better, but clumping up the extreme cases.
    
    \item \emph{Arbitrary (Fitted)}: Another possibility is a scale that is fitted arbitrarily. This can be a domain-specific definition from experts or a mathematical best fit with respect to some specific metric. With an arbitrary fit, it is possible to set the boundaries between distinct areas based on some criterion other than a mathematical distribution. The problem here is that such a fit could be highly subjective and only valid for a specific situation and point in time. One example is the energy labeling legislation in the EU \cite{european_commission_regulation_2017}: While in the past the distinctions from A (best) to G (worst) were sufficient, newer technology-enabled lower power consumption, therefore more categories have been introduced over time: A+, A++, A+++, and beginning with 2020 the scale will be completely rearranged to use the labels again A to G \cite{european_commission_regulation_2017}. Energy labels based on a quantitative ratio scale would not suffer this problem (e.g., labels with power consumption in Watt and efficiency in percent, or net and gross power labels). Further examples can be found in literature like Ho et al. \cite{ho_improving_2015}. They show and compare the arbitrary definitions of probability scales for words of estimative probability \cite{kent_words_2012}, defined in several different standards.
\end{itemize}

\emph{Decision 3: Assignment of Semi-Quantitative Numbers to Ranks}: Are the semi-quantitative assigned numbers centered around 0, increasing, or decreasing? Is the 0 included or not? Table~\ref{tab:scales} shows some examples of different scales, which are also visualized in Figure~\ref{fig:distributions}.

\begin{table}
\caption{Variants of classification schemes for the range from 0 to 100.}
\label{tab:scales}
\centering
\begin{tabular}{clllll}
Level & Linear & Logarithmic & Gaussian & Inv. Gaussian  \\
\hline
1 & $\quad0\dots20$ &  $\quad0\dots0.01$ & $\quad0\dots10$ & $\quad0\dots30$   \\
2 & $>20\dots40$    &  $>0.01\dots0.1$   & $>10\dots30$    & $>30\dots45$     \\
3 & $>40\dots60$    &  $>0.1\dots1$      & $>30\dots70$    & $>45\dots55$    \\
4 & $>60\dots80$    &  $>1\dots10$       & $>70\dots90$    & $>55\dots70$    \\
5 & $>80\dots100$   &  $>10\dots100$     & $>90\dots100$   & $>70\dots100$   \\
\hline
\end{tabular}
\end{table}

Aside from the distribution of values, the direction and location of the centers are significant \cite{thomas_risk_2013}. Humans are susceptible and biased towards different orders, and labels in scales \cite{moors_effect_2014}. Table~\ref{tab:scale-ranks} shows different variants of number assignments to ordinal scale levels.

\begin{itemize}
    \item \emph{Increasing}: Numbering the scale in increasing order. This relates to the notion of ``higher numbers result in higher risk''. 
    \item \emph{Decreasing}: Scaling the levels with decreasing ranges inverts the meaning. Here, less of something corresponds to higher risk, e.g., lower defense means a higher risk of successful attacks.
    \item \emph{Centered around 0}: Sometimes positive and negative aspects are modeled in the scale. e.g., losses or gains. In such scales, the neutral element is 0, while the more extreme cases fall to either side of the number range (positive or negative).
    \item \emph{Including or omitting 0}: If the scale includes 0 and the combination includes multiplication, this 0-level could completely wipe out all other properties, regardless of how high they are. This is unwanted behavior since it conflicts with the monotonicity and relevance criteria for coherent risk metrics \cite{artzner_coherent_1999}.
\end{itemize}

All these decisions are somewhat arbitrary and made mostly for convenience to have a more straightforward combination and ranking strategy later on.

\paragraph*{\textbf{(G) Range Compression}}
    Through pressing the real values into a scheme of ordinal scales, the original uncertainty ranges get lost, and the whole value range of an ordinal class is applied to the values. Overlapping ranges get clipped, smaller ranges get widened.
        
\paragraph*{\textbf{(H) Ambiguity}}
    The scales are often not defined precisely, and therefore can be argued and judged differently based on the experts' opinion.
    
    What is still light injury, what is already severe injury?
    Where is the border between once per week and once per month?
    How do \textit{``very low chance''} and \textit{``remote chance''} differ from each other?
    \cite{ho_improving_2015,windschitl_interpretation_1999}
    
\paragraph*{\textbf{(I) Neglecting Uncertainty}}
    By classifying, the original uncertainty in the judgment gets lost. The class imposes a new default range for the uncertainty. This relates strongly to range compression and quantification errors. If the uncertainty was huge and would span multiple classes, this cannot be encoded. If the uncertainty is smaller and would span only a small fraction of a class, this cannot be encoded either and gets lost in the process.
        
\paragraph*{\textbf{(J) Quantification Errors}}
    Especially on the border,  quantification errors can happen. If a value changes slightly, it could step up to the next level in the ordinal scale or fall to the lower level.
    This could change the result tremendously (imagine going from 2 to a 1 with a multiplied combination, resulting in half of the resulting risk, but in reality, the value just changed a little bit).
        
\paragraph*{\textbf{(K) Human Bias}}
    Humans are very biased \cite{ariely_predictably_2010,kahneman_subjective_1972,gilovich_heuristics_2002}. They are scared of bad outcomes and tend to underestimate the probabilities, or they are biased towards the other way and tend to overestimate bad outcomes (risk affinity bias). Also, humans tend to judge events based on their own experience, which is, by all means, also very flawed. Also, the cultural background and the daily condition play a huge role (see human inconsistency). This flaw can partly be covered by training for consistency and training for neutrality but is still there. Even if the probabilities are exactly defined, humans tend to misinterpret them \cite{windschitl_interpretation_1999}. Also, centering bias happens in this phase: Humans like to avoid the extreme values of a scale \cite{hubbard_failure_2009}.
        
\paragraph*{\textbf{(L) Human Inconsistency}}
    It is proven that humans are biased due to anchoring and framing. They change their judgments for the same questions based on the daily condition, the immediate situation before the judgment, or the scaling they have to do.

\subsection*{\textbf{Phase 3: Combining the ratings}}

After all the influence factors were rated according to the ordinal scales, they get combined. Most methods do this either multiplicative or additive, some have a weighting scheme for addition, and some also use a deduction factor for reducing the final score. All these methods have no mathematical foundation since ordinal scales do not define arithmetic operations (only ordering relation). Still doing it introduces many problems which are discussed here.

\paragraph*{\textbf{(M) Undefined Semi-Quantitative Arithmetics}}
As already mentioned, ordinal scales do not support operations like addition, subtraction, multiplication, or division. They only define an ordering relation for ranking them. Any semi-quantitative calculation with such ordinal scale levels is just an arbitrary approach, without any foundations or support from mathematics. Connecting to the example already discussed in problem (E): Ordinal scales - what is the meaning of multiplying two different classes of ratings? Can the words ``high risk'' and ``low severity'' be multiplied? No. However, we tend to believe that semi-quantitative numbers can. If a high risk corresponds to 3, and a low severity corresponds to 1, the result would have been $3\times1=3$. However, here we stepped over a fallacy because if we tried this with the corresponding textual labels, it is clear that this is an invalid operation (high risk $\times$ low severity = ?).
        
\begin{figure}
  \centering
  \includegraphics[width=\linewidth]{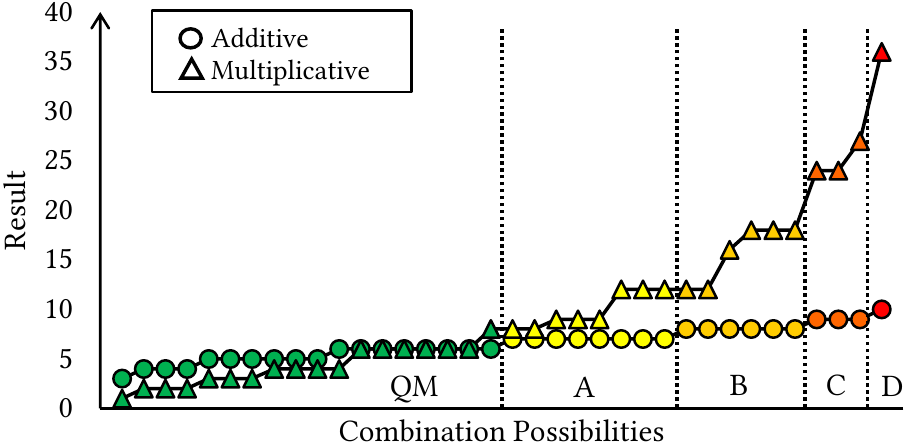}
  \caption{Comparison of additive and multiplicative combinations of values. The example is based on the Hazard and Risk Analysis, and the areas are the respective ASIL classifications defined in the ISO 26262 \cite{iso_26262}. While ASIL-C and ASIL-D are clearly distinguishable in both approaches, the boundaries of QM, A, and B are not that intuitive.}
  \label{fig:combination}
\end{figure}

\begin{figure}
  \centering
  \includegraphics[width=\linewidth]{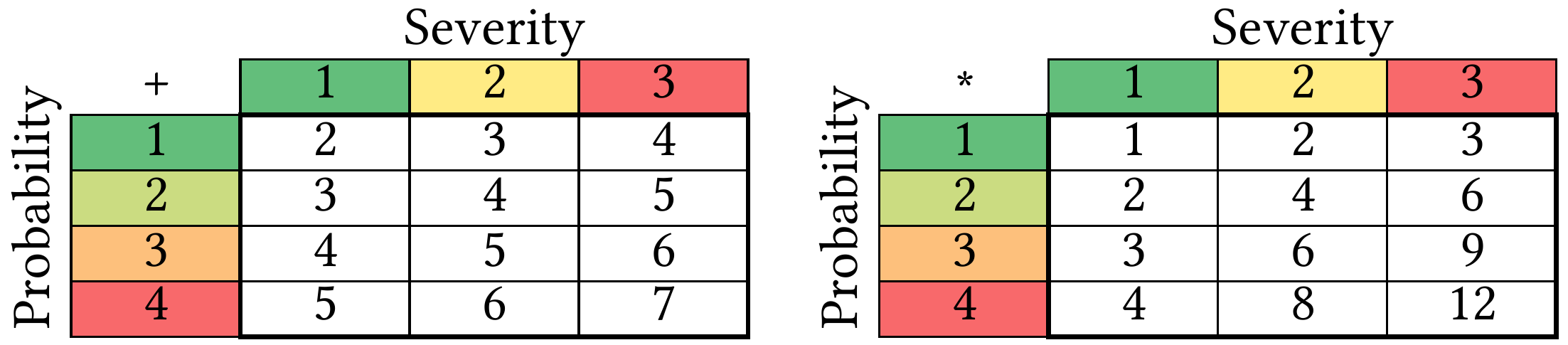}
  \caption{The tables show the results adding factors compared to multiplying them.}
  \label{fig:additivevsmultiplicative}
\end{figure}

\paragraph*{\textbf{(N) Arbitrary Combinations}}
The way ratings are combined invalid and undefined, but the actual operations are also chosen arbitrarily. Should we multiply or add up all ratings? Should an optional reduction be subtracted, or should the scale be inverted to have a more comfortable mathematical formula? Should we weigh the ratings before we add them together? How are the weights defined? This strongly depends upon the scale definition (see the problem (F)). On ordinal scales, there is no correct way to do this. It is just a convention or definition by some standard. Mostly the kind of combination is chosen to result in a nice number to judge the final risk. Figure~\ref{fig:combination} shows all additive combination possibilities of the HARA risk scores in the ISO 26262 \cite{iso_26262}, compared to the same scores when multiplied. By adding them, there are equal groups of levels used for further processing - but when multiplying, they do not group up that nicely, especially the border between categories QM, A, and B are not intuitively recognizable anymore. In addition to that, the table in Figure~\ref{fig:additivevsmultiplicative} shows numerical examples for the combination via addition and multiplication. We can see that multiplying produces more variance and also more risk classes than the addition.

Suppose the ratings include 0 as a number. In that case, multiplication could completely override all other ratings for a risk event, no matter how extreme they are (essentially making a risk irrelevant).

The definition of the weights and combination of influence factors for calculating the risks has developed into an industry because this is needed for actuaries, insurances, finance companies, and clinical and pharmaceutical industries. They try to define their weights and combinations according to some sophisticated model because, for them, it is the basis of million-dollar decisions. One example of a sophisticated combination is the algorithm for calculating the risk of unemployment in Austria's unemployment office \cite{ams-algorithmus}. It is a weighted additive scheme based on several demographic and social factors. The weights for combining the factors were inferred via historical data and statistics and reflected society's bias, which was highly disputed in the media. For example, female candidates had a higher risk of staying unemployed than male candidates, or that education was only a minor factor in getting a job.

\paragraph*{\textbf{(O) Dominating Components}}
If one property on the ordinal scale is low or high, it could dominate the others. For actual ratio scales, this is normal and reasonable. However, since ordinal scales lost their original real-world semantics and the numbers for the scale levels are just arbitrary definitions, the combination is not reasonable anymore. This could be a problem, especially if the scales have different distributions (one is linear, while the other is logarithmic), e.g., due to the ordinal scales, a level from a linear distribution gets the same influence as a level from a logarithmic distribution. Increasing or decreasing the linear level would result in linear response of the actual risk while increasing/decreasing the logarithmic level would change the risk by a factor of magnitude. However, the calculated ordinal risk metric would still change only by a linear term, regardless of the actual quantitative risk change. This contradicts the positive homogeneity, and the translation invariance property \cite{artzner_coherent_1999}.
    
\paragraph*{\textbf{(P) Neglection of Correlations}}
One of the most overlooked problems of risk matrices is the neglection of correlations. Cox et al. \cite{cox_whats_2008} stated that uncorrected negative correlations between risk matrices' influence factors could lead to worse than random results. While problem (B) already describes this, here, the actual effects are manifested. If we would know the correlations between the input factors, we could correct them in this phase by applying a correlation matrix or some other conversion factor to make up for this. This, of course, would only be possible when we had used quantitative risk assessments with ratio scales. On ordinal scales, it is difficult to model the effect one factor has on another. In the quantitative world, one can detect that a factor changes whenever another factor changes and how they are related to each other (positively or negatively correlated). Since ordinal scale levels are so coarse, this cannot be detected or corrected.
The consequences of this may be severe: Correlations could add up and result in a high-risk value, or they annihilate themselves, and the actual risk would not change even when the input factors change. All in all, the calculated risk metric does not reflect the changes in the real quantitative risk, which is a severe problem and contradicts the positive homogeneity property \cite{artzner_coherent_1999}.

\subsection*{\textbf{Phase 4: Ranking and Ordering the risks according to the resulting risk metric}}

In this phase, the combined risk scores get ranked again and ordered for their importance. As already was the case in phase 2, this ranking is again an ordinal scale and suffers from the same problems. Here, it is even worse because the source data is not a ratio scale but a combined ordinal score which drags along all the problems described until now.

In this phase, the combined risk score is again ranked on an ordinal scale, e.g., all values above a specific threshold get a high rank, all under a specific threshold a low rank, and all in between get medium. E.g., scores from 1 to 5 get a low ranking, scores from 6 to 10 get medium, and 11 to 15 get a high ranking. In addition to all discussed problems of ordinal scales, this phase has even more problems, partly since this final ranking is the basis for decision making.

\paragraph*{\textbf{(Q) Arbitrary Thresholds}}
The thresholds for the ranges of the final risk levels are often chosen completely arbitrarily, with a high emphasis on simplicity. In the hazard and risk analysis, for example, all scores until 6 are grouped to the lowest risk level (QM), and above 6, every whole integer represents an own risk level (7=ASIL A, 8=ASIL B, 9=ASIL C, 10=ASIL D) \cite{iso_26262}. This convention is convenient due to the combination method of addition. This makes it easy to estimate the final risk score already during the individual scores in phase 2.

In comparison, quantitative methods also define an arbitrary threshold in this phase, but this would consist of a distribution called the ``risk appetite'', which defines how much risk in terms of probabilities and real quantitative values is tolerable. For example, how much money loss is still tolerable with a 10\% probability, 50\% probability, or 90\% probability that the loss is realized.

\begin{figure}
  \centering
  \includegraphics[width=\linewidth]{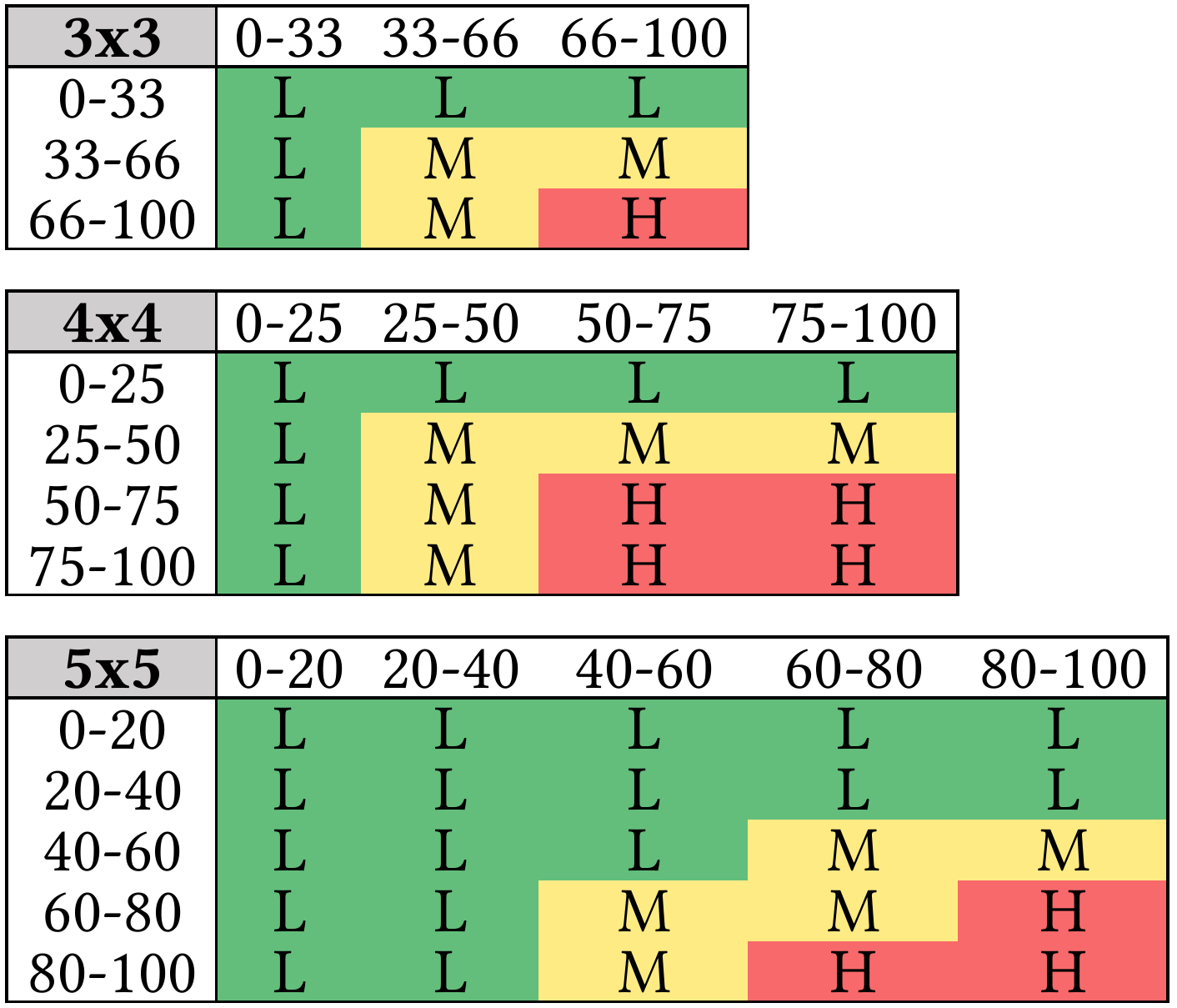}
  \caption{The only possible consistent assignment of risk matrices with linear input scales and a 3-rank output score (L=low, M=medium, H=high), for 3x3, 4x4, and one of the two possible colorings for 5x5, according to Cox \cite{cox_whats_2008}.}
  \label{fig:cox-consistency}
\end{figure}

\paragraph*{\textbf{(R) Inconsistency}}
In their work, Cox et al. \cite{cox_whats_2008} describe what consistency for a risk matrix means and why this is important to achieve that property. At the same time, they prove that full consistency cannot be achieved when ordinal scales are used. Consistency means that the resulting risk score should relate directly to the real quantitative risk. For example, it should not be possible to switch from the lowest risk category to the highest by doing just a small change during the evaluation. It should not be possible that actual higher risks get scored below actual lower risks.
Furthermore, all events in the same final risk class should represent the same level of actual risk, no matter how they are ranked, combined, and judged during the risk assessment. Cox defines three properties to ensure this: weak consistency, betweenness, and consistent coloring. Figure~\ref{fig:cox-consistency} shows examples for consistent risk matrices using linear scales as input and having 3 ranked ordinal scales as output (high, medium, low).

\begin{itemize}
    \item Weak Consistency: This property defines that all events which are in the lowest-ranked risk class should have lower actual risks than all the events ranked in the highest risk class. If these two classes are disjunctive, a risk matrix has at least weak consistency.
    \item Betweenness: It should not be possible to jump directly from the lowest risk class to the highest by just doing small changes in the input factors. Therefore, at least another class needed to create a border between the lowest and highest classes. This ``middle'' level may overlap with the lowest or highest class, but it should contain events higher than the lowest class and mostly lower than the highest risk class.
    \item Consistent Coloring: This property ensures that events on the same risk level should represent approximately the same actual risk in reality. It should not happen that two events are grouped into the same risk class but have opposite actual risks.
\end{itemize}

\paragraph*{\textbf{(S) Incoherence}}
Coherence is the notion of general properties for well-behaved risk metrics and was proposed by Artzner et al. \cite{artzner_coherent_1999} in 1999. They argue that a risk metric has to satisfy several properties (or axioms, as they called it) to be useful. Together these properties ensure that a risk metric is reasonable and well behaved. Since the final risk metric is obtained via a risk matrix, it should also satisfy these coherence properties:
\begin{itemize}
    \item Relevance: $X > 0 \implies p(X) > 0$\\
    When an event has an actual quantitative risk, the risk metric should also assign some positive value (the risk metric must not be zero). For ordinal scales which exclude the 0, this property holds.
    
    \item Monotonicity: $X\ge Y \implies p(X) \ge p(Y)$\\
    If a real event has a higher risk than another, the risk metric should also come up with a higher or at least the same values. This is already sometimes violated due to the classification into ordinal scales. Using ordinal scales, an event with lower risk might get a higher score than an event with actual higher risk. Just recall the example of oil leakage from the motivation section.

    \item Translation Invariance: $ p(X + \alpha r) = p(X) - \alpha$
    This means that, by making some additional effort to reduce the risk, the respective risk metric should decrease by a corresponding amount. It should not happen that increasing or decreasing a risk produces an incoherent change of the risk metric. This also implies that if some action reduces multiple risks by the same amount, their relative order to each other must not change.
    
    \item Subaddidivity: $p(X+Y) \le p(X) + p(Y)$
    When combining two risk events, the risk metric should be at most the addition of the single risk metric values. If the events overlap or are somehow correlated, it is less than the sum of the individual values.
    
    \item Positive homogeneity: $p(\lambda X) = \lambda p(X)$
    This property ensures that the risk metric reflects affine changes in risk. If the risk doubles, also the metric should double.
\end{itemize}

Real quantitative methods would support these properties already with the most basic risk equation: $\mv{Risk} = \mv{Impact} \times \mv{Probability}$. This equation fulfills all the mentioned properties of coherence and consistency when used with ratio scales or probability distributions.
        
\paragraph*{\textbf{(T) Ambiguous Order}}
    If the risk matrix is at least weakly consistent, the highest and lowest-ranked risks can be ordered, but what about ordering inside the classes? If different risks result in having the same score, they cannot be prioritized anymore.
    Furthermore, the middle classes may partially have the same quantitative risk as the lowest or highest classes, making the ordering not very intuitive. An event with a middle-classed score may have a higher actual risk than the higher class score. Also, due to range compression, the highest risks get all clumped up together in the highest class, but the differences could be orders of magnitudes apart.
        
\paragraph*{\textbf{(U) Risk Inversion}}
    The problem of risk inversion is a very severe one. Lower risks might get a higher score than actually higher risk or vice versa.
    We will repeat the thought experiment from the motivation section again to make this clear:
    
    Think of an assessment of environmental contamination. Two means of transportation are compared: a car which leaks half a liter of oil every week, and a plane which happens to leak 100 liters every half a year.
    Furthermore, imagine the following scale for oil leakage:
    \begin{itemize}
        \item $0\dots0.1$ liter = Low impact (1)
        \item $0.1\dots1$ liter = Medium impact (2)
        \item $1\dots10$ liters = High impact (3)
        \item $>10$ liters = Very high impact (4)
    \end{itemize}
    \noindent
    and the following time scales for the frequency:
    \begin{itemize}
        \item[(4)] Daily = Very high frequency (4)
        \item[(3)] Weekly = High frequency (3)
        \item[(2)] Monthly = Medium frequency (2)
        \item[(1)] Yearly = Low frequency (1)
    \end{itemize}
    
    The car would get a medium impact (2) and high frequency (3), which would result in a risk score of 6, but leaks about 100 liters per year. The plane would get very high impact (4), but only a low frequency (1), which results in a risk score of 4, and the quantitative leakage is 200 liter per year, which is double the leakage of the car. This is a typical example of risk inversion, where an actual higher risk event occurs to be scored with a lower risk score. Even if the plane's frequency rating would have been medium (2), the score would still only be a little higher than the one of the car, while it should be double as high if it really would represent the actual risk.
        
\paragraph*{\textbf{(V) Unknown Uncertainty and Confidence}}
    We already established that neglecting the uncertainty in phase 1 was not a good idea. Here we have to pay the price for this. Range compression and quantification errors have additionally contributed to completely wipe out any notion of uncertainty or confidence in our data. We cannot determine the uncertainty in our data anymore and are left only with the given ranges coming from the scale levels through the risk scores alone. Maybe our estimations have been very uncertain and may, therefore, be wrong? On the other hand, if we were very confident in our estimations, the scales' ranges and the arbitrary calculations increased the error and uncertainty. How can we ever know if we neglected them along the way?
    
    Quantitative methods could accomplish this by propagating the uncertainty throughout all calculations, or even better, by using the values' distributions to consider even more details of the underlying quantitative data.
    
\subsection*{\textbf{Phase 5: Making decisions based on the Risk Assessment}}

Based on the scoring and ranking of the risks, we want to decide which ones we want to mitigate and which ones we can tolerate. This last step strongly relates to decision theory \cite{howard_foundations_2015}. Here, human bias plays a huge role again: the simple traffic-light system of a risk matrix is very appealing for management people. Also, just the task of talking about risks already gives an impression of achievement and benefit. Nevertheless, Hubbard et al. \cite{hubbard_how_2016} debate that this impression may be deceitful and is just a perceived impression, not a real one. Simply discussing risks may already induce some satisfaction and the notion of accomplishment, but, as Hubbard argues, to make sure that the methods are beneficial, we have to measure their performance. Unfortunately there are little to no evidence that qualitative risk matrices work \cite{julian_talbot_whats_2018}, and even that is just a pure argumentative one, but there are many pieces of evidence that they have quite some problems \cite{cox_whats_2008,cox_risk_2009,thomas_risk_2013,hubbard_how_2016,hubbard_how_2014}. One big problem regarding the measurement of performance is that there could be years until some risk eventually occurs - hence there is no immediate feedback, which could be measured easily.

\paragraph*{\textbf{(W)   Wrong Impression of Benefits}}
    Because the risk assessment based on semi-quantitative methods seems so ``easy'' and ``natural'', there is the notion that it is correct and trustworthy. Risk matrices are established tools, which companies have used for many decades now. They may even appear to be ``authoritative, and intellectually rigorous'' \cite{thomas_risk_2013} due to their seemingly correct semi-quantitative approach. However, as we established in this work, this is not the case. The benefit could be just an illusion, again bred by the human bias of uncertainty aversion and authority bias \cite{kahneman_subjective_1972}.

\paragraph*{\textbf{(X) Deferred Feedback}}
    Hubbard et al. \cite{hubbard_how_2014,hubbard_how_2016}, and others \cite{newport_deep_2016}, stated the actual fact, that immediate feedback is an absolute must for being able to improve. The longer the feedback loop endures, the weaker the learning effect is. For risk assessment methods, the time frame between the assessment and the actual risk event may be years apart. Therefore, the initial evaluation is seldom reviewed for correctness, and methods themselves are even more rarely approved for their validity or performance. Often, the people who did the assessment already left the company long before, making it even more difficult to reevaluate and improve on the estimations. Unfortunately, this is also a problem that applies even to quantitative methods. It is important to check the validity of methods by measuring their prediction strength and comparing this with other methods to find the most suitable method for a purpose.
\section{Conclusion}
In this work, we discussed many aspects and problems of risk matrices. We showed that in every phase of the risk assessment process, risk matrices have flaws and may introduce errors that could lead to wrong decisions in the end. By showing this, we made another case against qualitative or semi-quantitative risk assessment methods and proposed quantitative approaches. In our research group, we are currently developing such a method based on quantitative risk assessment for cyber-security, called RISKEE \cite{krisper_riskee:_2019}. The mathematics behind it can be used for any risk assessment, and we will make it available as soon as we have enough evidence supporting the correctness. In the future, we plan to investigate problems that exist even when using quantitative methods, e.g., detecting incompleteness or irrelevance of input factors and tackling the problem of deferred feedback to evaluate the appropriateness of the method. Also, combining several different expert judgments to get a realistic judgment is an area we want to tackle in future papers.

Our plea is to the safety and risk experts out there to reflect on the possible pitfalls of risk matrices and review their methods and estimations, whether they may have fallen into some of the possible traps luring inside risk matrices. Furthermore, we encourage using quantitative risk assessment methods wherever possible.

\bibliographystyle{IEEEtran}
\bibliography{90-references}

\end{document}